# On the design of a profession-oriented course on Theoretical Mechanics for physics education students

**Marianne Korner**[1,*] **and Christos N. Likos**[2,#]

***ORCID https://orcid.org/0000-0003-2117-0482**

**#ORCID https://orcid.org/0000-0003-3550-4834**

[1] *University of Vienna, Faculty of Physics, Experimental Basic Training and University Didactics, Porzellangasse 4, 1090 Vienna, Austria*
[2] *University of Vienna, Faculty of Physics, Computational and Soft Matter Physics, Kolingasse 14-16, 1090 Vienna, Austria*

**ABSTRACT.** We report on a profession-oriented course we offered at the University of Vienna, aimed at physics education teacher students. The course on Theoretical Classical Mechanics has been conceived and designed from its outset with the explicit goal of bridging the gap between the abstract, mathematical notions employed in Theoretical Physics with the concrete future needs of prospective teachers in their profession. We aimed at countering both the negative attitudes of students towards Theoretical Physics and the interrelated skepticism of professors regarding the students' mathematical proficiency. Our main findings are that these goals can indeed be achieved through a careful selection of course material and the associated mathematical tools, by closely interwoven lecture topics and exercises, and thorough planning according to principles for high-school teaching known from science education research. Establishing close connections between the material taught in the course and the students' future occupation as high-school teachers has proven to be of utmost importance. This is possible without any sacrifice of mathematical rigor or of the quality of Physics presented.

## I. INTRODUCTION

The present study deals with our approach towards a class on Theoretical Physics 1 (Classical Mechanics) for physics education students at the university of Vienna, and it is located within the tradition of Action Research.

The underlying historical background of this lecture on Theoretical Physics 1 (T1) is that in past it has been a major hurdle in the education of pre-service teacher students, as the first in the row of a total of four lectures on theoretical physics. As such, it plays a key role not only in and of itself but also due to its pivotal character in introducing the prospective teachers into the tools, ways of thinking and abstractions associated with theoretical physics in general. Despite its key character for the whole curriculum, it is an unpopular lecture in the students' perception, an issue that puts the necessity of motivating the students' interest and engagement with the course in particularly emphatic terms. It has been frequently reported by students that they cannot see a sense in learning things about which they perceive that they will never need them in their professional life as high school teachers. Nevertheless, mechanics is a key topic for teachers and, moreover, it is the topic which comes closest from all aforementioned lectures on theoretical physics to the demands in high school teaching. Beyond that, as a teacher, one needs to command a much deeper level of knowledge in a certain field than the level at which one teaches. Out of the many reasons justifying this statement we mention here two: First, to be able to address the vast diversity of questions posed by the high school students in class; second, to be able to justify certain decisions on the selection of contents and on the level of elementarization within the Model of Educational Reconstruction [1].

As part of a negative spiral, which we term “vicious circle” in what follows, the students’ lack of motivation and negative perception about the T1 feeds into the professors’ attitude towards the audience, so that it also becomes an unpopular lecture to teach. Thus, often lecturers complain that students “do not know anything” and “do not learn anything”, which is a statement that rather indicates frustration and can lead to resignation.

Moreover, this vicious circle leads in more cases than desired to high dropout rates in the exams. Therefore, we, a team of two researchers had the idea to improve the lecture, not by lowering the professional standards but by improving motivation of the students and by a thorough planning of the content.

The authors of this article undertook the task of jointly designing and offering this *course* during the summer semesters 2022 and 2023 at the University of Vienna. In doing so, they combined their expertise in high school teaching and science education research on the one hand (M.K.) and in research and teaching in the field of theoretical and computational physics on the other (C.N.L.). This work is meant to describe the process of designing a lecture suitable for physics education students, revealing the underlying considerations and bringing together the expertise of persons working in different fields, which is a novelty. In this work, we present concrete indications that our approach has been successful instead of a strict empirical validation, which will be topic of future work.

## II. CONTEXT OF THE STUDY

### A. Pre-service teacher programs in Austria

The overall population of Austria is about 9 million and there are only a handful of universities offering study programs in Physics Education. The University of Vienna with its 90 thousand students is not only the largest in Austria but also one of the largest in Europe. It is thus not a surprise that indeed as many as 90 % of the currently employed roughly 1,500 high-school Physics teachers nationwide are graduates of the University of Vienna.

In Austria as well as all over the world, lecturers have to deal with requirements of a diverse and heterogeneous audience. Therefore, our findings are relevant for all lecturers who aim to treat a non-uniform audience best. In our case, we identified two major conditions for teaching physics at university level which we considered.

Firstly, we identified a broad span of different levels of mathematic skills and practice. In our case one of the reasons for this lies in the fact, that prospective teachers have to choose two different subjects that they wish to study; in contrast to other countries, there are no restrictions regarding the combinations of these subjects. What can be seen as freedom of choice can also lead to unusual or - depending on the point of view - unfavorable subject-combinations. One of the popular combinations with physics is mathematics, having the advantage that the students are more used to or comfortable with mathematical formalisms compared to students who combine physics with, e.g., any subject in the Humanities. On the other hand, for the practice of in-service teaching, uncommon combinations of subjects may lead to an enhanced job satisfaction, a deeper understanding of learning hurdles and to more creativity in choosing appropriate and interesting contexts on the basis of which content is conveyed [2, 3].

On the basis of our personal inquiry in several groups of physics education students, and according to statistical reports of the University of Vienna [4], we found that approximately only half of all students study mathematics as a subject combined with physics. Thus, for teaching physics at university, we as lecturers have to take into account that we face a group of students covering a broad span of different levels of mathematic skills and practice. We cannot rely on an external mathematical training beyond our lectures and exercises; therefore, the design of physics classes has to consider this fact and has to be planned accordingly.

Secondly, at the University of Vienna as well as at some other universities, two different bachelor physics degrees are offered, which barely have an intersection: Let us call them *physics education* and the *scientific discipline of physics*[1]. Therefore, a certain physics content – in our case mechanics – must be taught within different curricula. Depending on the curriculum, a different number of week-hours is assigned to each course. The selection of content, intensity, and depth in concept must also be tailored to the amount of time available. Seen from this point of view, it becomes clear that a direct comparison of the physics-related skills and knowledge of different groups is inappropriate. For us as lecturers, this implies the duty to think carefully about what should be done, what can be done, and what goals should be pursued for the physics education students in this short time. On the other hand, the goals for the two bachelor's degrees are different. Physics education students aim to teach at schools, with a more or less clear picture in mind how their profession will look like. Therefore, lecturers are well-advised to plan their teaching according to what physics education students are going to need in their professional future. This is what we call a *profession-oriented course*.

### B. Lectures and *courses* in university teaching

Traditional lectures are a controversially discussed teaching format. However, the curriculum of the University of Vienna prescribes this format for the T1, hence there is no choice about that. In the following, we want to share a few thoughts about lectures and show possibilities of conducting a lecture in a way that initiates and supports best learning processes in the students.

Historically seen, the origins of university teaching can be seen as triangulated from: *lectio - quaestio – disputatio*, where the necessary and intended interactions of all involved partners facilitate academic education [5]. Nevertheless, compared to other teaching formats, university didactics offers lecturers little theoretically sound and practically relevant help in planning and delivering lectures and the empirical findings about the efficiency of lectures are anything but clear [6, p. 94].

It is evident that a lecture can be seen as a direct instruction. The utmost challenge is to combine this with elements that foster the learners' active engagement in the subject matter in the sense of constructivist learning theories [7]. This is also supported by Hake's meta-study [8], conducted with N = 6542 participants of introductory physics courses. He reports that students attending courses that make use of interactive-engagement methods achieve deeper conceptual understanding than those in traditional courses. Employing the Force Concept Inventory [9] or the Halloun-Hestenes Mechanics Diagnostic test [10], Hake's study reveals a difference between the groups with and without interactive-engagement methods of approximately two standard deviations in average normalized gains. Moreover, students in interactive courses show better abilities in problem solving strategies.

Based on a review on literature, Reinmann [6] offers several options for the design of a lecture in order to provide opportunities for an active construction of knowledge~~, which are discussed in the~~

---

[1] While the bachelor studies of *physics education* students comprise subject A, subject B, and fundamentals of Educational Science, the students of the *scientific discipline of physics* spend their whole time on physics. Thus, for the former group approximately 42 % per semester are dedicated to physics as opposed to 100 % per semester for their peers in the scientific discipline.

~~following~~: Lecturers can modify the monologic communication style of a lecture by implementing phases of group work or group discussion. One can add different forms of representation, e.g., use visualizations or demonstrations or, also in theoretical physics, conduct experiments. Additionally, one has the options of choosing the content to be covered, using different argumentation patterns and selecting illustrative examples thoroughly. On the other hand, lecturers can encourage students to use different learning strategies and remind them in regular intervals of the purpose of the lecture.

To some degree the authors consider some of the above-mentioned points included in and underpinned by the work of a science education researcher, Muckenfuß [11], who introduces the term *meaningful contexts*. Meaningful contexts on the one hand support the learners' interests (answer to the question: *Why should the learn that?*), and on the other hand are selected in such a way that the imparting of the specific content structure of a discipline is optimally supported. By adopting these considerations for our planning, we aim at transferring findings concerning high school teaching to university-level teaching.

An additional relevant finding from science education research, from which parts can be considered for the planning of a lecture, is Collin's approach called *Cognitive Apprenticeship* (CAS) [12, 13]. It represents one of the most interesting realizations of a moderate constructivist learning environment and distinguishes between easily accessible object knowledge and merely implicit expert knowledge. Another work regarding design issues for learning environments by the same author [14] claims that the first issue to address concerns what he labels *authenticity of a learning environment*: providing an answer to the question for the potential use of the knowledge and how to plan a learning environment accordingly. He states that students have to be prepared for "the kind of complex tasks that occur in life" (ibid, p. 4) and avoid an acquisition of knowledge without the ability to apply it. In our perception, a course should satisfy the demands of the subject matter as well as those of teaching and learning seen from a moderate constructivist perspective, connecting thereby both issues pertaining to the students' future profession of teaching at the high school. For the design of such a course on T1, we purposefully aimed at transferring the results from science education research at high school level to university teaching.

## III: METHODOLOGY

In this section we outline what the aim of Action Research (AR) is, where do we see ourselves located within this multi-layered construct, and how we can use it for our purpose to improve the T1 lecture.

Kurt Levin is perceived to be the inventor of the term Action Research (AR) which he described as a spiral process of steps to be carried out, namely: planning – acting – observing – evaluating [15]. Along the years, there have developed different kinds of AR, which all share the feature as to abstain from common quantitative and even qualitative research approaches. Moreover, Kemmis, McTaggart, and Nixon [16] identify two common features attributed to all kinds of AR: It either deals with the recognition of the capacity of a certain group involved in a research process or it focusses on making a certain improvement in practice, as in our case.

As long as there is a wide variety approaches to AR, we concentrate on the three kinds of AR as described by Carr and Kemmis [17]: technical, practical, and critical AR. Technical AR is guided by an interest in improving control over outcomes, critical AR by the aim to emancipate people from an unfair situation, and practical action research guided by an interest in "practice-changing practice" [18].

*Practical AR* adopts a focus we can identify best with. Greenwood and Levin underpin this when they outline the aim of AR to be "…social research carried out by a team that encompasses a professional action researcher and the members of an organization, community, or network ("stakeholders") who are seeking to improve the participants' situation" [19, p. 3]. In accordance with Feldmann [20, p. 242], the aim of our actions is to critically reflect on our own practice in order to improve it and gain a better understanding of the mechanisms behind it. As researcher-practitioners we aim to act reasonable, sensitive to the students' needs to promote their performance in the exam and their mindset regarding comprehension of (theoretical) physics.

As long as the term AR can be perceived in different ways, we orientate on five principles in order to frame our work and justify it as AR [21, p. 13-14]. We are directly concerned with the situation being researched (1) and we start from very obvious and practical questions (2) to improve the situation of teaching and learning T1 within the demands of the faculty and the curriculum (3) with manageable effort for all participating persons (4). To a certain degree we intend to increase social justice (4), as long as the learning aims of the lecture should be feasible also for students who do not have mathematics as a joint study. Finally, putting together this paper and keeping in mind that there will be further opportunities for teaching T1 stands for the continuing effort to improve its quality and hopefully other lectures (5).

Concerning the data collection, we rely on notes we took in the course of our discussions with former lecturers, students and while explaining time and again our own perceptions of the situation to each other. Moreover, while running the third cycle as described below (Figure 1), we integrated the continuous feedback of an exercise instructor and of our student tutor. Together, we had during our weekly meetings a number of discussions comparing and reconciling our perceptions. In these exchanges we deliberated the selection of exercise tasks; about the difficulties and questions the students had in the exercise sessions and the tutorial; the justification of aspects of the content matter in the lecture; and the available options to support the students' learning processes. We analyzed and discussed the feedback we received from the students during and after the lectures. Additionally, we put particular emphasis on both of us attending the lecture and the exercises. Alternating, one of us did the teaching and the other person observed the students' actions, perceived to a certain degree their mood and – most important – was able to take notes on these issues. Finally, we incorporated into our analysis the reports, automatically created by the learning platform used (Moodle) about the course activities, and the feedback we received from the evaluation questionnaire of the university. For the design of our AR-process we oriented on the cycles of planning (Figure 1) described by Riel [22]. The following section is structured accordingly.

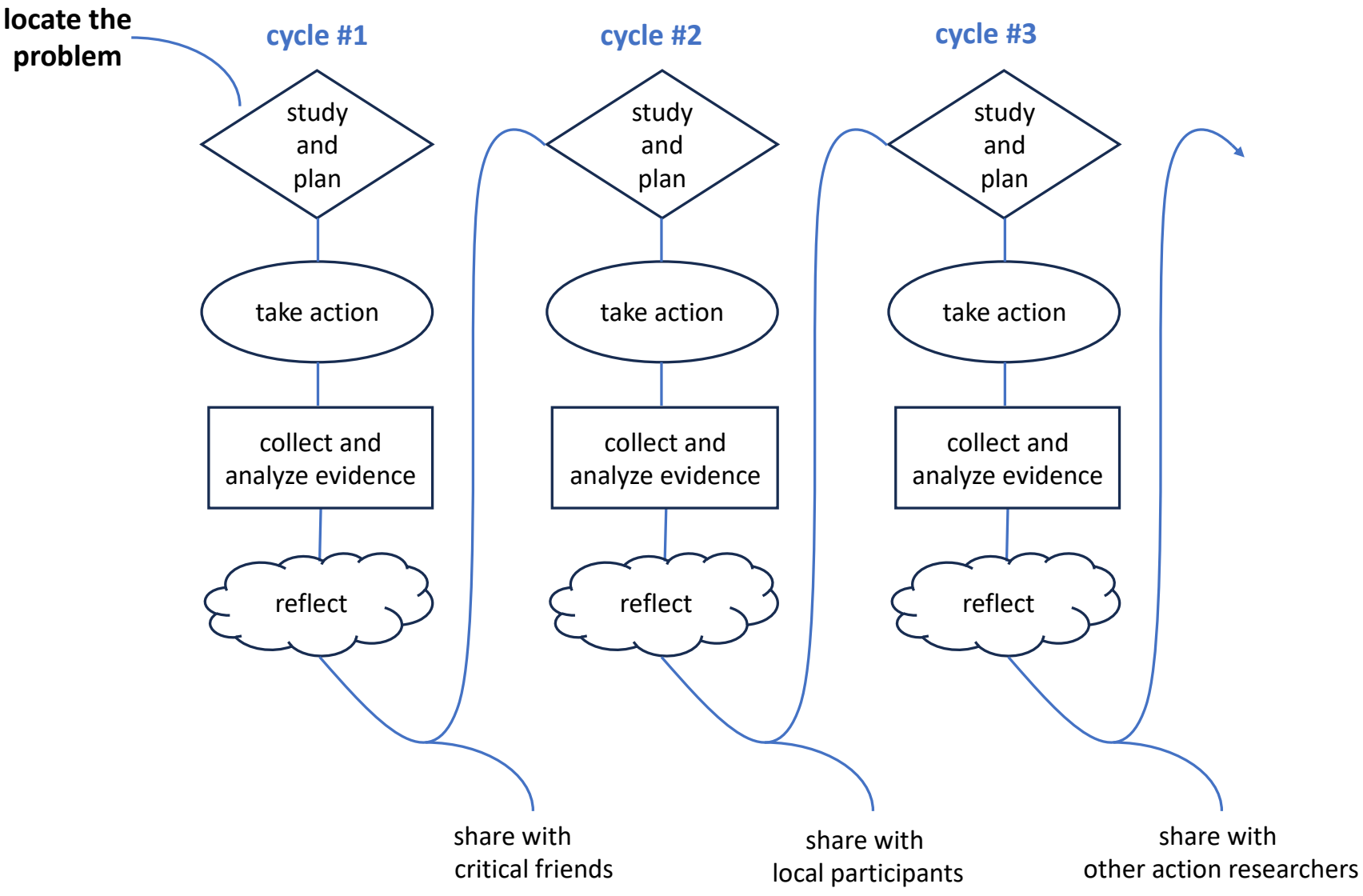

*Figure 1: Cycles of planning. Graphic according to Riel (2003)*

## IV. PROCESS AND FINDINGS

### A. Cycle #1: Identifying the problem and reflecting on our own special abilities

First of all, we identified the problem. For this purpose, we talked to all parties involved: former lecturers, students and tutors, and listened to their experiences and points of view. We had various exchanges with lecturers who were not happy with their students' effort, their progress and their performance in the final exams. We heard the well-known statement, reporting that certain contents "had been explained dozens of times" and found that they were still "not understood."

We let students tell us about lecturers who seemed not to meet their interests, learning requirements and expected needs in their later profession as high school teachers. We talked to exercise instructors of the corresponding exercise sessions who attested to the students a fundamental lack in mathematical skills. In return, we took a step back and again talked to students and listened closely how they felt about the exercise sessions. Finally, as a kind of communication link we let tutors, who were more experienced students of higher semester, tell us about their perception of the students' and previous lecturers' points of view. All in all, a confusing and incoherent picture emerged from the puzzle pieces of the conversations. The only common element was that the vast majority of the stakeholders was dissatisfied.

At this point we, the two authors, reflected on our special skills: On the one hand there is Marianne Korner, who has been a physics teacher at high school for nearly three decades and parallel to this started a professional career in science education research and holds now a position as an educator for pre-service teachers at the University of Vienna. Based on the experiences in these two areas, input can be given on both: what student teachers need in their future profession and on how to teach according to constructivist ideas of learning in the best possible way. One the other hand there is Christos Likos, Professor for Computational and Theoretical Physics at the University of Vienna, who has a long-standing experience in teaching Theoretical Physics Courses for students of the scientific discipline of physics (as labelled on p. 6) and practices the subject in his research as well. Likos' previous experience with prospective high school physics teachers has only taken place in a mixed context with regular physics students, during his tenure at German universities, where the distinction between the two study programs is, depending on the Federal State, less sharp than in Austria. Both of us, perceived as experts in our special fields, decided to bring together our respective fields of expertise on an equal footing, with the purpose of designing a *course* on Classical Theoretical Mechanics as described above.

Throughout our discussions and involvement in the matter, the main questions we were interested in, and which guided our AR crystalized into be the following:

- How can T1 be made meaningful for teacher students?
- How is it possible to incorporate constructivist elements into a lecture?
- What possibilities can be identified to positively influence the students' outcome in the exam?

### B. Cycle #2: Planning a course on T1 applying the Model of Educational Instruction

The starting point of our considerations was the above-mentioned statement, according to which the (former) lecturers did everything in their capacities as to make Classical Mechanics intelligible to their students, falling nevertheless short of achieving the desired outcomes on the learning progression. We interpret this fact as a failed cognitivist approach to teaching and learning, best described by the teaching-learning shortcut, as put forward for the first time by Holzkamp [23]: the inadequate idea that things that have been *taught,* are concurrently *learned*. In order to avoid this, we oriented at constructivist teaching and learning approaches, which have been demonstrated to work better [e.g. 24] at least under certain circumstances, as to take learners' pre-instructional conceptions into consideration [25].

A process containing a multitude of discussions between the two authors started, according to which content issues and issues on science education research were presented, weighed, and negotiated. Looking back, this led to an enormous learning process on both sides, which is documented in this section. To begin with, we applied routines known from science education research for the planning of teaching at high school level on the teaching at university level. For this purpose, we employed a model commonly used for planning sequences of lectures in science education: the Model of Educational Reconstruction (MER) [1], see Figure 2.

(3)
**Design and evaluation of teaching and learning environments**
Issues of real teaching & learning environment are taken into account

(1)
**Clarification and analysis of science content**
Subject matter clarification
&
Analysis of educational significance

(2)
**Research on teaching & learning**
Perspectives of the learners
(conceptions & affective variables)
Teaching & learning processes
Teachers' views & conceptions

*Figure 2: The Model of Educational Reconstruction - MER, figure redrawn after [1, p. 21]*

The MER consists of three components. Component (1) contains the clarification and analysis of the content structure seen from an educational point of view. According to this, one has to analyze what the specific concepts are (subject matter clarification) and has to take the aims of teaching into account (analysis of educational significance). This process means that lecturers need to make certain choices within the curriculum, regarding both the content matter and the aims of teaching.

One of the key ideas of the MER implies that domain-specific knowledge has to be transformed into knowledge suitable for educational purposes. This transformation involves the conversation of the content structure of a particular domain into a structure for instruction, as schematically shown in Figure 3; notably, these two structures are significantly different. Therefore, science content structured for a specific topic cannot be directly transferred to an instructional context. In a first step it needs to be *elementarized* to make it accessible for learners, which means it should be broken up into tiny pieces, which jointly form a certain concept. We will call these pieces the *key ideas* of a concept, in accordance to Paul Hewitt [26]. In a subsequent, second step, out of these key ideas a content structure suitable for instruction has to be built. Within this process, both issues of content matter and students' perspectives on it should be at best considered; this is expressed by component (2), Figure 2. There already exists research on teaching and learning, which offers us a great deal of insight into students' conceptions as well as affective variables such as interest, self-concepts and attitudes toward learning. Both components, (1) and (2), have a mutual influence on each other as indicated by the two small horizontal arrows in Figure 2.

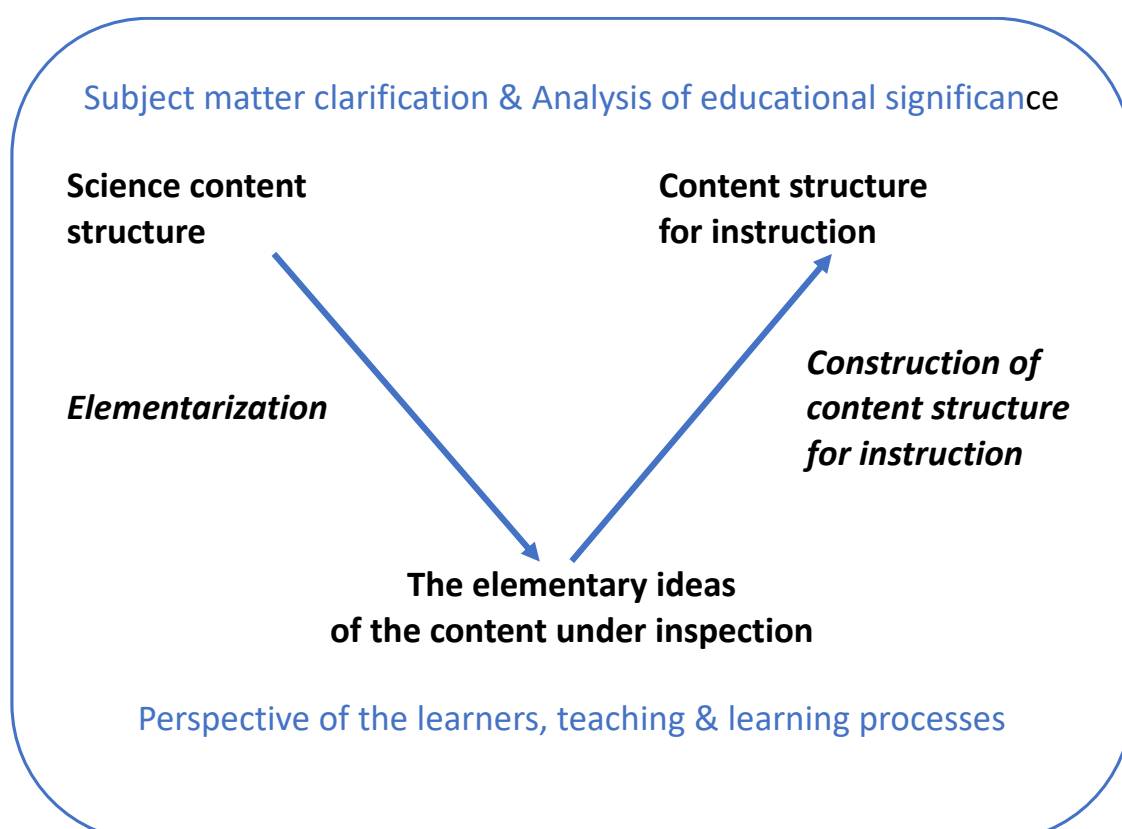

*Figure 3: From subject matter towards teaching [1, p. 21]*

Finally, component (3) outlines the design of the actual teaching environment, which is structured according to the students' needs on the one hand and on the outcomes of the restructuring of the domain-specific knowledge according to educational issues.

In the following section we present the outcomes of our considerations applying the MER model (components 1&2) to our course on Theoretical Physics 1 (Classical Mechanics).

*Component (1) of the MER: towards a content structure for instruction* --The curriculum of the University of Vienna [27] for Theoretical Physics 1 for teacher training students in the bachelor (subsidiary 4th term, 2 ECTS) comprises the following content matter:

> *Newton's laws as differential equations, Kepler's problem, harmonic oscillator, reference systems, Galilean transformations, conservation of energy, principle of least action and Lagrange formalism, symmetries and conservation laws, Hamilton formalism.*

Within these given topics we have considerable freedom in the emphasis, depth and precise content and in the choice of the specific tools employed to transmit content to the students. Interestingly, the motion of rigid bodies does not form part of the official curriculum, which is in our view a serious drawback of the latter that we attempted to redress in designing our course.

As indicated above, a very significant part of the students' frustration with T1, and the accompanying perception of the lecturers' regarding the students' abilities and motivation, stems from the lack of sufficient mathematical preparation of the former, which on the one hand hinders their progress in adopting certain concepts of mechanics and on the other places students with mathematics as a joint study into an undue advantage with respect to their peers. To address these issues, we identified the crucial areas of mathematical competencies necessary for conceptual progress in T1. Furthermore, we designed strategies to help students overcome the inhibiting mindset that mathematical abilities are a fixed, pre-determined talent as opposed to a set of skills that one can develop by adopting a positive attitude and by exercising repeated practice.

Table 1displays a list of **mathematical skills** we focused on and we restricted ourselves to. In addition, we provide a summary of the reasons behind our decision to put particular emphasis on these.

*Table 1: List of mathematical skills we focused on and the reasons behind our decisions.*

| # | **Mathematical content** | **Reasoning** |
|---|---|---|
| 1 | Complex numbers | Their significance for the harmonic oscillator (with damping/driven) cannot be overemphasized; their |
| | | necessity for the later lecture on quantum mechanics is evident. |
| 2 | Polar coordinates | Essential for the Kepler problem |
| 3 | Vectors and their representations (notation in a line) | Emphasizing the geometrical nature of vectors and thus a distinction between a vector and the way it is described by different observers; underlying the difference between objects and their representations. |
| 4 | Making use of the orthogonality of vectors (inner product) | Essential for the concepts of work and power. |
| 5 | Cross-product of vectors and its interpretation in various contexts | Essential for the concepts of torque and angular momentum |
| 6 | Ordinary differential equations (ODE) of second order and their transformations through change of variables | ODEs are needed both for the harmonic oscillator topics and for the Kepler problem, let alone for the whole area of Lagrangian Mechanics. |

Concerning the physical content matter, which is determined by the curriculum, we focused on three main contexts: the harmonic oscillator, Kepler 's problem, and inertial forces. Each of these problems was described in one of our "three languages" in mechanics, the Newtonian, the Lagrangian, and the Hamiltonian, with emphasis on the first two.

Our starting point was always a *problem* to be solved. By doing so, we pursued two main goals: firstly, we focused on the students' conceptual development in physics, and secondly, we used physics as a *meaningful context* in the sense of Muckenfuß [11] for teaching the above-mentioned mathematical skills. All in all, the planning comprised two tiers of contexts: within the given curricular requirements we thoroughly chose meaningful contexts (*problems*) for the physical concepts to be learned, and these examples worked as contexts for the mathematical skills to be developed.

*Guiding principles for the implementation of the course on T1* – Accomplishing the detailed planning of the course, we consented to various useful of strategies, which became our guiding principles throughout the semester.

- First of all, and repeatedly during the lectures, we emphasized on the fact that physics as a discipline describes the outcomes of experiments; even theoretical physicists rely on experiments, and whatever one calculates has to correspond with reality or with a real event in nature. It is always a legitimate argument in physics to state "it is so, because experiment says so".
- Building upon the idea of *one reality – many points of view* offers an intuitive way to explain the necessity of inertial forces: "What is a straight line for one observer is a curved one for another." Similarly, one and the same vector can have different "names" and different rates of change depending on the choices of various observers.
- Concerning the Lagrangian, we avoided all unnecessary abstraction and circumnavigated all proofs, by presenting this formalism as solution technique to address problems that are otherwise unsolvable due to constraint forces. Regarding the Hamiltonian formalism, we introduced the canonical momenta by drawing analogies with the Legendre transformations in thermodynamics. The idea behind these decisions is that once the students see how well the formalisms work, they become convinced about their importance and usefulness, acting thereby against the attitude "I don't need this for my future work."

- We straightly focused on the description of nature in terms of physical quantities that can be directly perceived (force, torque, acceleration) and thus avoided abstract but (in this context) useless theoretical concepts such as the Principle of Least Action or Poisson brackets. These tend to reinforce the (false) impression that theoretical physics is too removed from practical reality and therefore uninteresting.

*Component (2) of the MER* – Considerations according to component (2) of the MER focus on the perspectives of learners and the teaching and learning process, whereby the latter is discussed in detail within Cycle #3.

According to the perspective of the learners, a main issue was to impart knowledge on the relevance of the course content on their future profession as physics teachers in high schools, thus providing a profession-oriented course. To meet these student interests, we aimed to repeatedly demonstrate which of the concepts of the lecture were applicable to high-school teaching. Moreover, we to linked this knowledge to empirically tested teaching concepts for high school [28], which on the one hand are addressed in subsequent lectures within the physics education studies and on the other hand are part of the curricula for high-school teaching.

Special focus was laid on the learners' previous experiences by considering the prior knowledge of students:

- What do students bring from their own high school education?
- What was taught in previous classes at university?
- What are the mathematics and physics skills of the students? e.g.: How about different notations?

We will unveil these considerations by the following example: at the very beginning, we addressed vectors, covering the key ideas that a vector is not identical with its representation and that nature needs no coordinates. To illustrate this and frame it with an obvious context, we chose the nose of Boltzmann's bust, which is in a niche of our lecture hall. The nose exists, and so does the location vector, which links nose and an observer, independently of who observes and how we describe it. As an example, different coordinate systems were considered, and in the following even different types of coordinates, namely Cartesian, and polar ones. One emphasis was put on the representation of a vector (Cartesian coordinates as well as polars) as a sum with explicit display of the coordinate unit vectors, instead a column. We learned from teachers in high school and from lecturers in the introductory lectures in physics that students tend to be more familiar with vectors in columns. Therefore, we explicitly took into account that learning about the representation of vectors as sums with explicit reference to the unit vectors is a learning step that students have to undergo and planned the explanations and exercises accordingly. In a broader context, emphasizing the nature of vectors as geometrical objects facilitated both the understanding of inertial forces and analogies between non-inertial frames and geometrical shapes with intrinsic curvature.

Summarizing, we planned our course considering the partially conflicting aspects involving time, aims, and pre-instructional requirements. Additionally, we especially focused on students whose second subject was other than mathematics.

As a preliminary finding from the considerations described in Cycle #2, we strongly experienced that educational and content issues go hand in hand. In our opinion, one will not succeed if considering and planning one without the other.

### C. Cycle #3: Implementation (component 3 of the MER)

According to literature [6], and relying on the outcomes of science education research concerning constructivist learning environments [7] we implemented a variety of means fostering the students' communication among themselves and with us lecturers at best.

- well-structured lecture with short breaks
- discussions on *context-oriented situations* rather than fragmented artificial tasks
- giving room and encouraging students repeatedly to ask (clarifying) questions
- tasks and phases of group work during the lecture, e.g. to work out mathematically demanding calculations and to think them over or to discuss solutions;
  aim: technical skills indispensable for the analysis of an important problem, no exam material
- fostering a culture of discussion by allowing students to express diverging opinions regarding the outcomes of thought experiments or their interpretation
- implementation of professional learning groups (PLGs) via Moodle
- tutorial
- weekly self-assessment questions (guided repetition of the content of the lecture) – intended as a basis of discussion in the tutorial
- bonus-tasks (as preparation and scored as bonus points for the exam) – upload via Moodle; scored by tutor and/or lecturer
- close collaboration with the exercise instructors
- tasks in the exercises time-coordinated with the content of the lecture
- weekly meeting of the team: lecturers, exercise instructors, tutor
- complementarity between university didactics, science education research and the subject sciences within phases of team-teaching
- "*bridging*": Repeatedly supporting the perception that this course has an impact on further teaching and thus becoming meaningful for students as prospective teachers by linking …
  - … students' pre-instructional experiences from school (e.g. concerning notation, prejudices regarding inertial forces) with the course's content
  - … the course's content with possible questions from high school students, and
  - … the content of the course to (well-) known and empirically tested teaching concepts [28] from science education research, known from parallel lectures
- Q&A session during the Easter break
- timely answers to mails
- exam insight

In the following, we outline a specific example to illustrate phases of group working within the lecture. The context chosen was to describe the movement of a (small) mass in a central potential (Kepler's problem), e.g. to derive the parametric equation of the trajectory $r(\varphi)$ and characterize possible orbits, which turns out to be a considerably tricky problem. To deal with issues 1 and 6 from Table 1 (complex numbers, ODE), we followed the strategy of breaking down the solution of the ODEs into smaller sub-tasks. We let students work in small groups to reconstruct the necessary steps to bring the ODE from the original form to a new one and to derive the solution. With previous knowledge of the harmonic oscillator ODE, it turns out to be at the same time astonishing and didactic that the Kepler problem's ODE becomes identical to the harmonic oscillator with a suitable choice of variables. As a guidance, a document on the learning platform (Moodle) was provided which gave hints and broke up the task into elementary steps.

As a link to the students' future work at school, we provided an animation with GeoGebra, a mathematical program which combines dynamic geometry, algebra, and calculus and which is commonly used at Austrian schools [29]. On the one hand, ready-to-use applications are available, and on the other hand GeoGebra can be employed to visualize and investigate a large variety of mathematical problems. In Figure 4 is shown an application visualizing conic sections [30], as used during the lecture. Figure 5 displays a possible trajectory of a satellite/planet, which is described by the equation of motion derived in the lecture. Designing this simulation was one of the tasks of the corresponding exercise sheet, which students had to complete on their own.

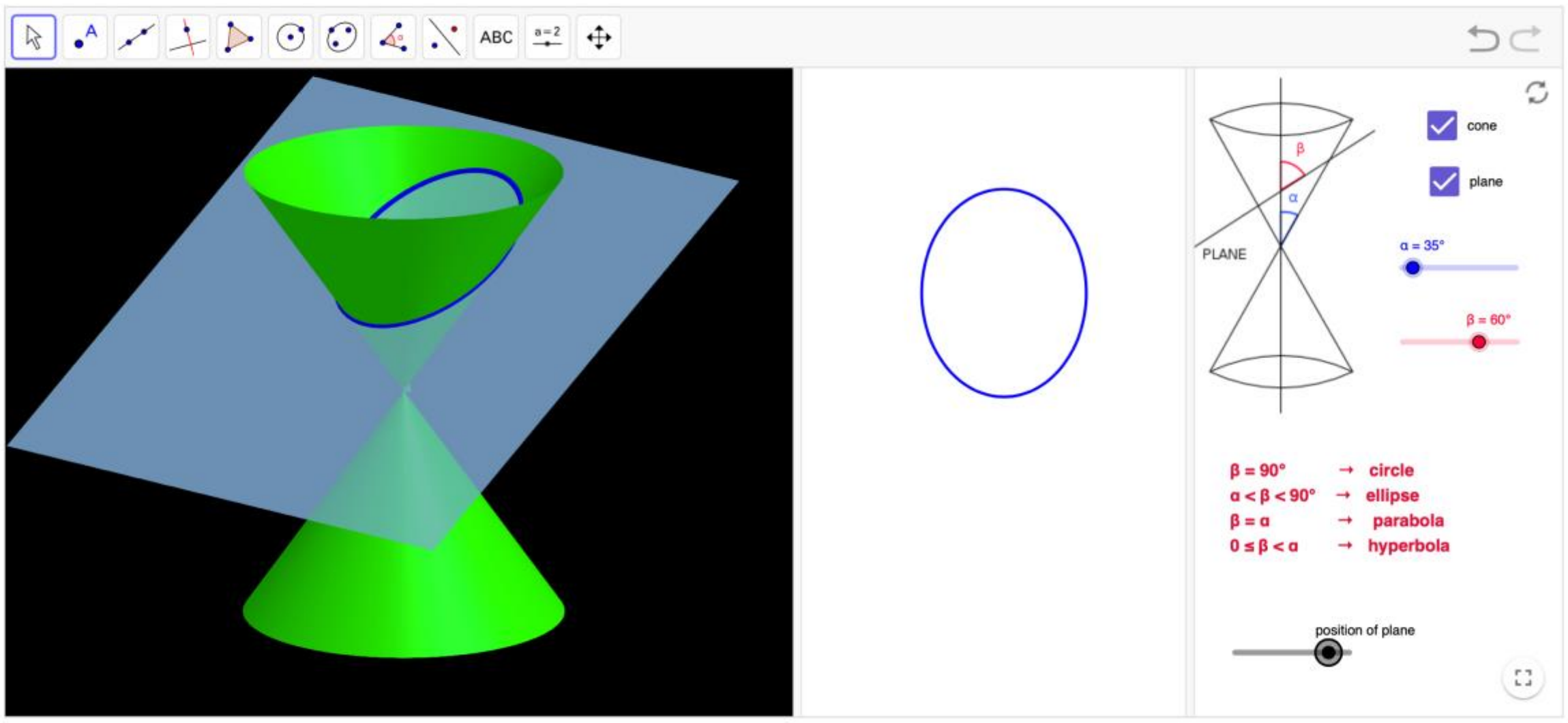

*Figure 4: Screenshot, visualizing the conic sections with GeoGebra. [30]*

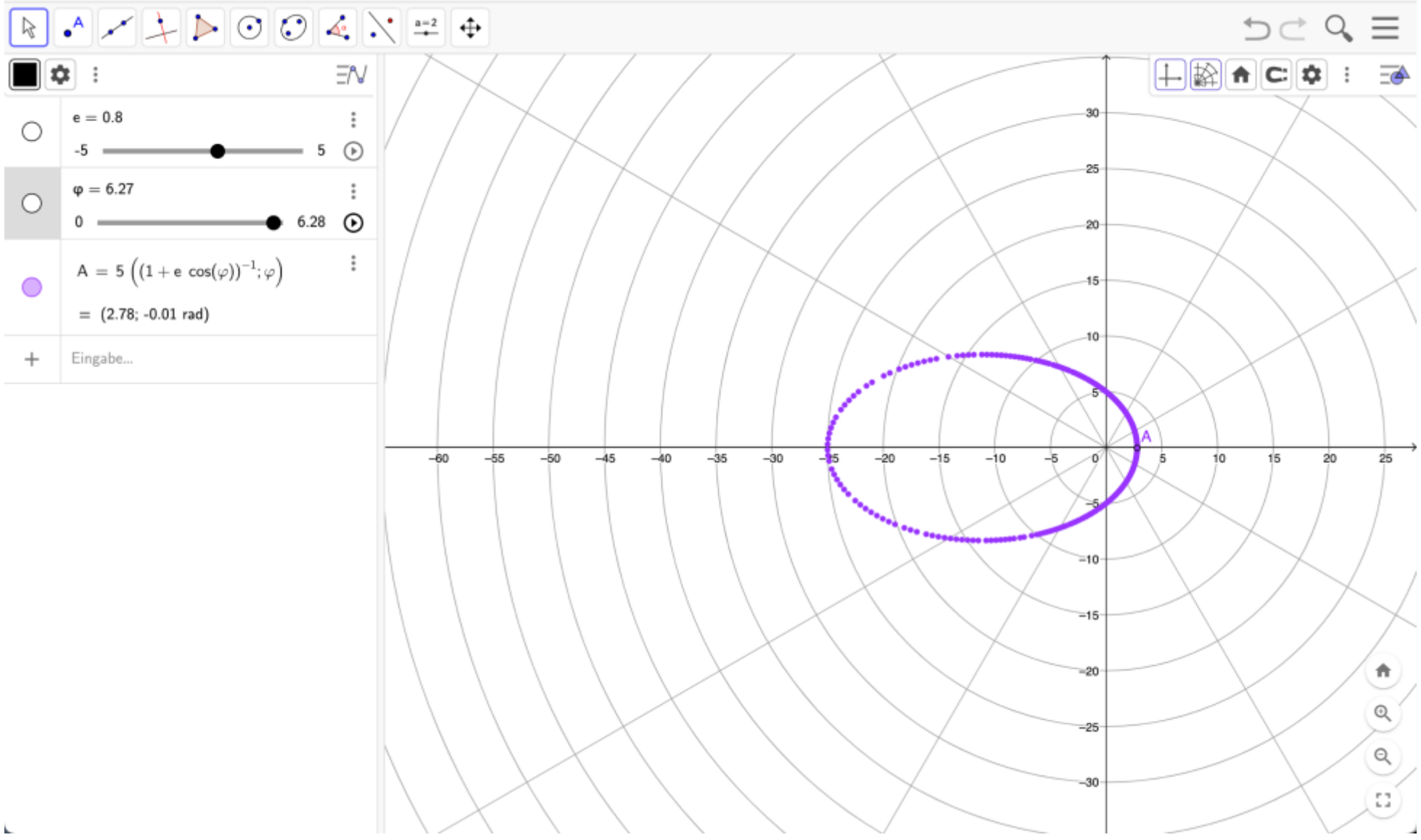

*Figure 5: GeoGebra screenshot of a graphical representation of the trajectory in a gravitational potential: slider controller for the eccentricity e of the resulting ellipse is chosen, but fixed; slider controller for the angle $\varphi$ runs from 0 to $2\pi$.*

Both simulations have in common that there are slider controllers provided, by which parameters can be varied, either by choosing a fixed value of a variable or by letting them run across a chosen interval. Figure 5 demonstrates both cases: The eccentricity is fixed at the value $e = 0.8$ is while the polar angle $\varphi$ runs across the interval $[0; 2\pi]$ creating the purple curve (ellipse).

## D. Findings (cycle #3)

### *Quantitative data*

While the first two cycles mainly concern considerations made in advance of the course, cycle 3 is dedicated to the implementation of the course on T1. We therefore report at this point about selected results of the implementation in the summer semesters 2022 and 2023.

The total number of students enrolled in the lecture in summer semester 2022 as well as in 2023 was N = 89 in both courses, offering therefore the opportunity to compare absolute numbers.

To get a first impression of students' engagement, let us look at the activities on the learning platform (Moodle) associated to the course (Figure 6). On the vertical axis the number of accesses during the semester is counted, which gives a rough impression about the activities, but not a deep insight about how close the activity is linked to cognitive processes.

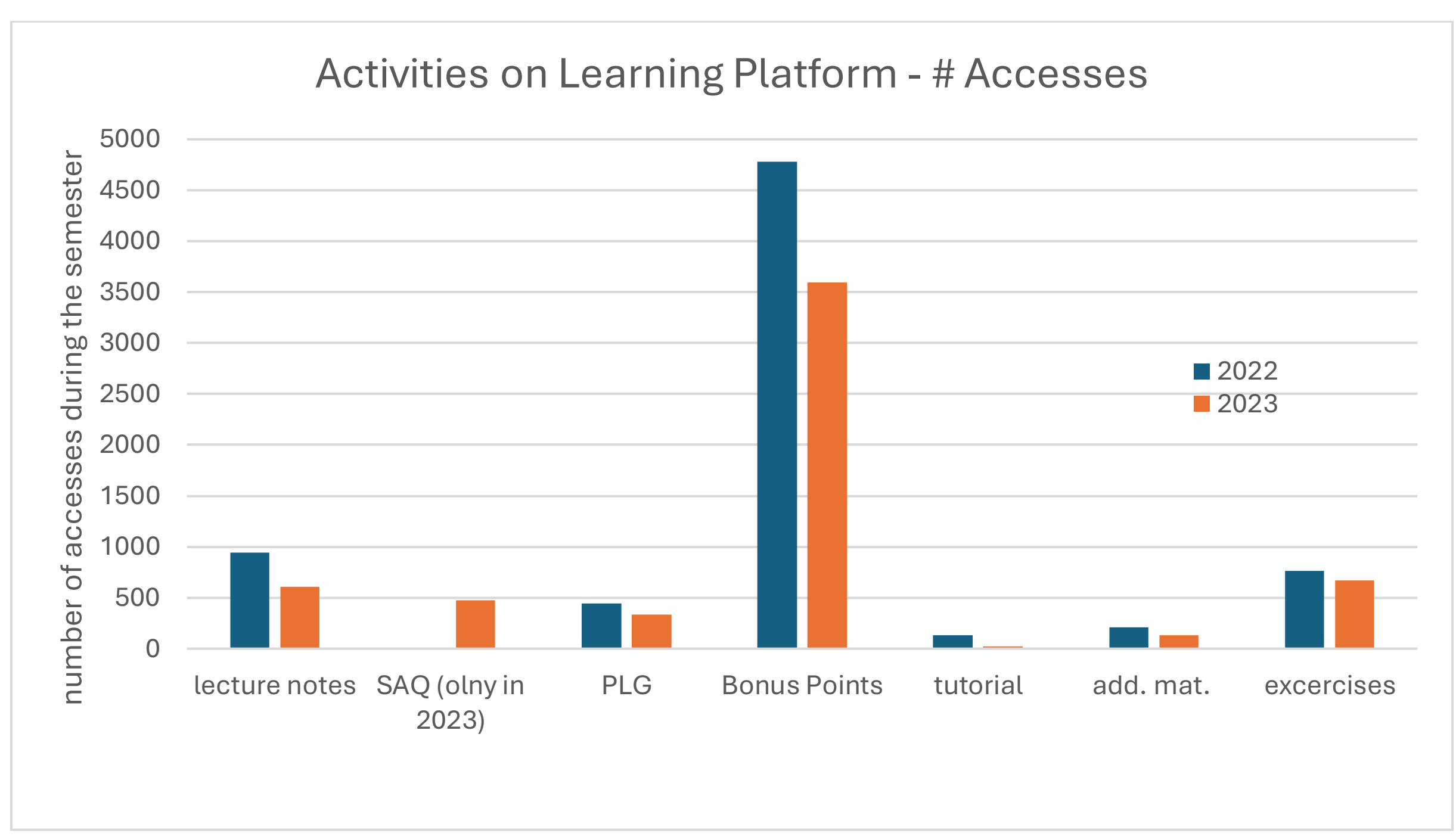

*Figure 6: Overview about the students' activities on the learning platform. SAQ = self-assessment questions, which were not available in summer term 2022; PLG = professional learning groups; bonus points for the exam; accompanying tutorial: only questions before the final exam; add. mat. = additional materials concerning Kepler's problem; exercises: tasks and solutions for those students who did not enroll in the exercise-course.*

A deeper insight is offered by the mean number (plus standard deviation) of students engaging in the various activities offered on Moodle. In Figure 7, we report exclusively on those sets of data for which a comparison is meaningful. In general, we can conclude that from 89 students enrolled in the lecture about one third engaged regularly into those activities, which were meant to support the students' learning.

Trying to identify trends within these data, it becomes evident that the engagement into the different activities was in 2023 lower than in 2022. This interpretation is underpinned by the observations during the tutorial: the tutor reported that in 2022 about four times as many people regularly took part in the tutorial as in 2023.

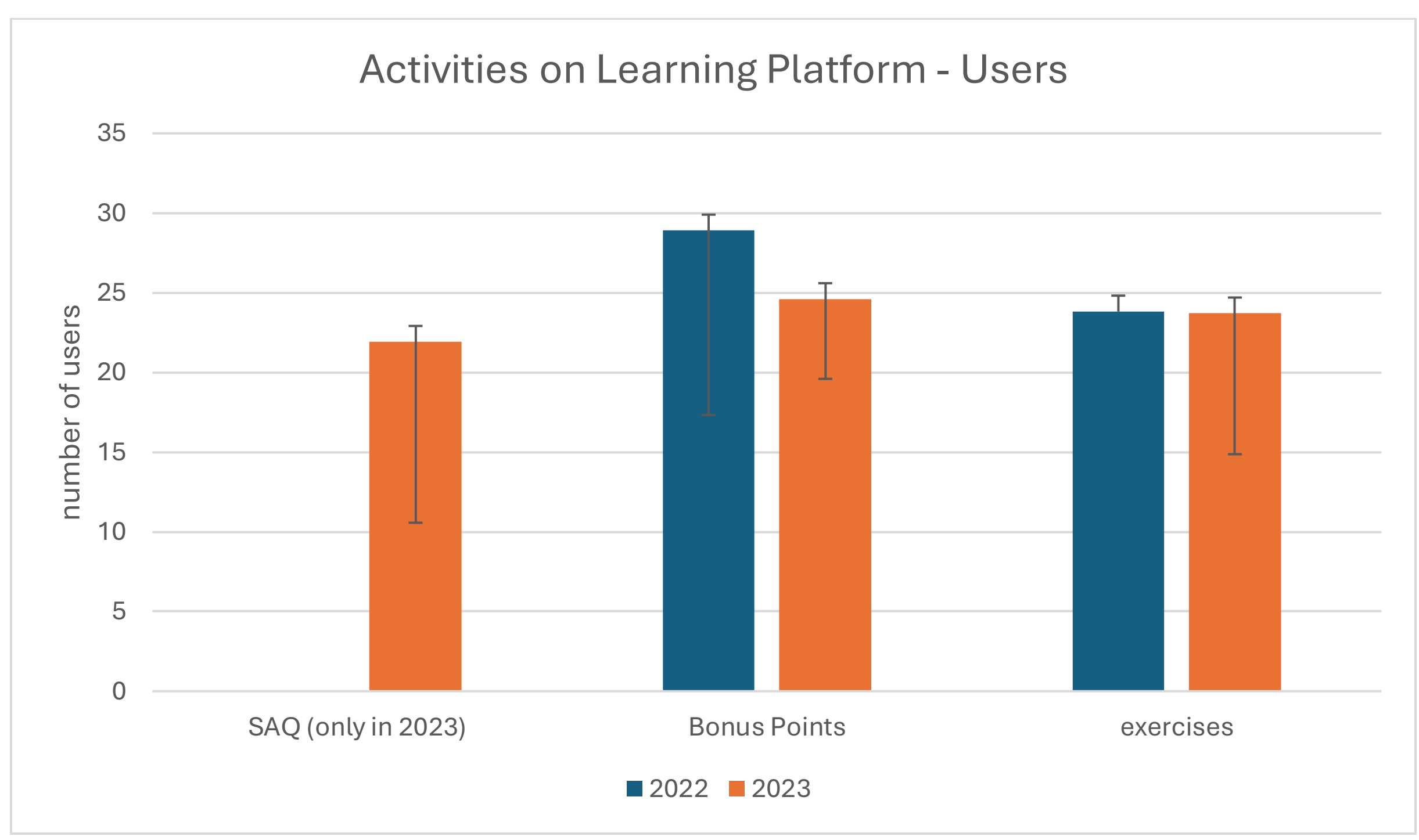

*Figure 7: Analysis on the number of users participating in the different activities on the learning platform. SAQ = self-assessment questions, which were not available in summer term 2022; BP = bonus points for the exam; exercises: tasks and solutions for those students who did not enroll in the exercise-course.*

In the second half of the semester, we carried out an evaluation of the course, using a standardized questionnaire provided by the University of Vienna. We use the part with the open-ended questions as an additional corpus of analysis. The two questions we focus on are:

1. What is particularly good about the course, what is not (e.g. pace, amount of material, requirements, expected prerequisites, etc.)? What suggestions for improvement do you have?
2. If there was a tutorial: Comments on the tutorial (How did it support your learning success? Feedback for the tutor)

Positive feedback concerned mainly the possibility of bonus points; knowledgeable, motivating and friendly lecturers; pleasant working atmosphere; fast response to feedback; clear explanations; often indicating the relevance of the topic for teachers; intermingling of subject-specific and pedagogical aspects. In 2022 some students gave positive feedback for the tutorial, and they emphasized its necessity for passing the exam. In contrast, in 2023 there have been only two replies concerning the tutorial which stated that the students had not used it. Negative feedback complained about too fast pace during parts of the lecture and difficulties to write and think simultaneously; lacking previous mathematical skills of the students; in 2022 a student wrote that the exercises have been too difficult.

We now contrast the engagement into the Moodle activities and the feedback with the outcomes of the exams. For each round of out T1-course we are obliged to offer 4 examination dates over which the sum is calculated. Figure 8 displays the data which show that in summer semester 2022 the exam activity ($N_{total.\ 2022}$ = 32 out of 89 beginners) was slightly higher than in summer semester 2023 ($N_{total,\ 2023}$ = 28 out of 89 beginners). These data match our counting during the lectures, revealing that in 2022 on average 30 to 35 students attended the class, where it was 20 to 25 in 2023. Moreover, the percentage of students who passed the exam was 68.8 % in 2022 and 53.6 % in 2023 out of those students who took the exam. Referring to all students enrolled in the course, the percentages are 40.0 % (2022) and 31.5 % (2023).

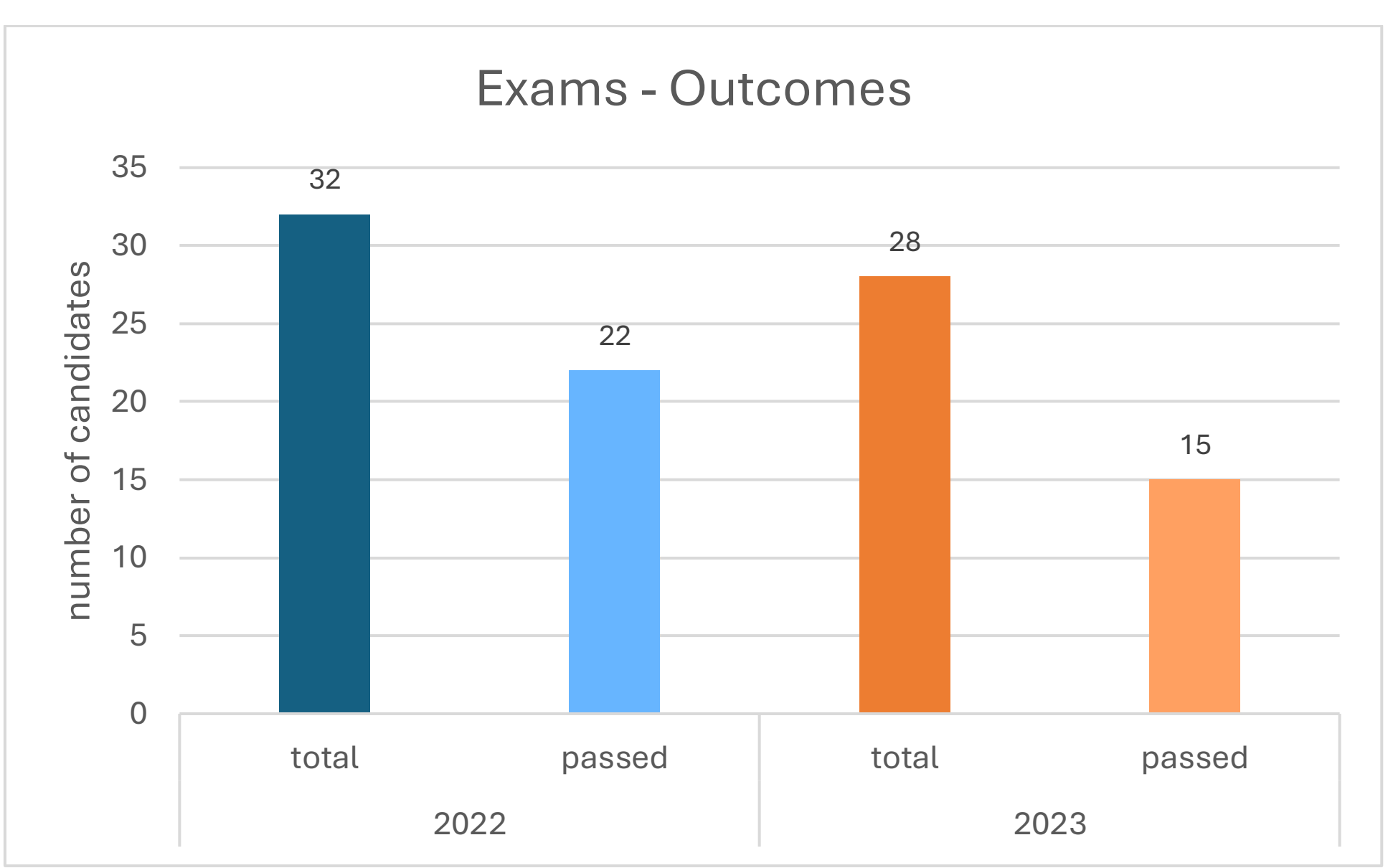

*Figure 8: Total number of students taking the exam vs. number of students who passed the exam (for both rounds of the T1-course).*

***Qualitative findings***

In the second half of the semesters, the course was evaluated, employing a standardized questionnaire set up by the University of Vienna. We use the part with the open-ended questions as an additional corpus of analysis. The two questions we focus on are:

1. What is particularly good about the course, what is not (e.g., pace, amount of material, requirements, expected prerequisites, etc.)? What suggestions for improvement do you have?
2. If there was a tutorial: Comments on the tutorial (How did it support your learning success? Feedback for the tutor)

Positive feedback obtained concerned mainly the possibility of bonus points; motivating and friendly lecturers; pleasant working atmosphere; fast response to feedback; clear explanations; frequent comments by the lecturers indicating the relevance of the topic for teachers; and intermingling of subject-specific and pedagogical aspects. In 2022, some students gave positive feedback for the tutorial, and they emphasized its necessity for passing the exam. In contrast, in 2023 there have been only two replies concerning the tutorial which stated that the students had not used it. Negative feedback included complaints about a too fast pace during parts of the lecture and difficulties to write and think simultaneously; lacking previous mathematical skills of the students; and in 2022 a student wrote that the exercises have been too difficult.

To complete our data collection, we carried out N = 8 interviews with students who attended the course and took the exam (4 each in 2022 and 2023). The main parts of the semi-structured interviews covered two aspects: first, the impact of peer activities on student learning, and second, whether the students had become aware of the relevance of the specific contexts we focused on.  An additional open question concerned the perceived difference to other lectures.

The possibility of completing the bonus exercises (bonus points) was reported most of all and it was highly appreciated, followed by the implementation of PLGs. There are few lectures with comparable features. Most of the other lectures do not foster student peer activities, so that students rely on their own networking activities, mostly by WhatsApp. The reported advantage lies in the structured way in which student activities have been guided by these questions, the voluntary and stress-free engagement in physics contents, and the fact that feedback was provided.

Concerning the profession-oriented feature of the course, it was appreciated that there have been at several points cross-references to the students' future teaching at schools, as one of the lecturers (MK) contributed her expertise from high-school teaching. This led to the perception that the course on T1 was closer to the knowledge needed for teaching at school. Although there have not been additional suggestions on how to improve the course, some interviewees claimed that the amount of time spent for completing all tasks was rather high compared to the credits given.

The student feedback given on the specific contexts we focused on was diverse. Our favorite, Boltzmann´s nose, was hardly mentioned, whereas the activities and explanations around the inertial forces, which is an important as well as a confusing topic, were highly appreciated. The clear differentiation between inertial and rotating frames, explained with reference to contexts of daily life such as an accelerating streetcar or a merry-go-round, led to conceptual insights that were not present in the introductory lecture on experimental physics. Our choice to introduce and focus on polar coordinates was valued because a deeper, quantitative insight into the planetary trajectories was reported.

We finally inquired about our decision to use GeoGebra within the course. GeoGebra was perceived as a useful means for visualization, specially making use of the slider controllers in that instrument. Particularly appreciated was the fact that using this tool does not put students with mathematics as a second subject in a privileged position, since everyone but a single student was familiar with it from their school. Digital tools were reported to be employed rarely at university level, in contrast to high school teaching and therefore our course was said to be a positive exception.

The open-ended part of comments on our course revealed that the reference and the linking of mathematics to experimental facts was rated positively. Mathematics did not appear in the course as serving its own purpose but rather as a useful and effective tool towards the description of physical reality. Three students emphasized that we have been the only lecturers who managed to foster an atmosphere of discussion about concepts (and thus conceptual understanding) and who treated students as discussion partners. By our interviewees with study experience from 4 to 8 semesters, our course was rated as one of the best.

## V. DISCUSSION AND IMPLICATIONS

We first start with the lecturers' impressions and contrast them to the findings reported above. We as lecturers were somewhat surprised that out of 89 students enrolled in each course, only about one third (summer semester 2022), resp. one fourth (summer semester 2023) attended the lectures constantly. Approximately the same percentages we find for the exercises. Therefore, the number of students who are actively taking exams is considered rather low. This drop-out problem concerns a lot of different lectures and is not specific to T1. Some of these students even quit their studies within the first semesters.

If we look at the percentage of students who passed the exams referring to those who took it, we can draw a better picture, especially for 2022 where 68.8 % passed it. Moreover, for the first examination date in 2022 about two thirds of the students passed with the two best grades, 1 and 2. We explain this with the students' strong commitment to all activities in and around the lecture (bonus points, tutorial, open question sessions, questions during the lecture) and the positive mood we perceived when communicating with them. In 2023, the students did not engage in these activities to the same degree (see Figure 6 and Figure 7) *and* performed below the results of 2022 in the exam. The results of the T1-exams of 2022 and 2023 match outcomes reported by other lecturers we talked to: The cohort which attended T1 in 2023 performed within other lectures and seminars below the average of the last years and was perceived as less active. At this point, we assume a strong interdependence between these two findings. Unfortunately, there exist no reliable data available on how the students performed in the years before. Nevertheless, we know on the one hand that peer activities have not been previously implemented in the way we did. Moreover, a comparison of our own courses from 2022 and 2023, shows that more student engagement improved the outcomes in the exam.

Nevertheless, after the first round of the T1 course in 2022, we even tried to enhance the activities to guide student learning by implementing self-assessment questions (SAQs) to each lecture. The underlying intention was to facilitate repetition of the content with a focus on conceptual understanding within the tutorial. According to constructivist learning theories, the SAQs should provide a guidance for students through the subject matter, facilitating self-organized learning. Moreover, the implementation of PLGs (professional learning groups) was intended to support the students' exchange with one another and thus the social character of learning, which is another feature of constructivist learning settings. We have no data on the extent to which the students made use of the PLGs, but for the SAQs the tutor reported that the tutorial was not used in this sense by the majority of students. Therefore, for the upcoming third round of the T1 course we will put even more emphasis during the lecture on the meaningfulness to engage into these questions.

Linking the data of students' activity to the outcomes in the exams, it seems that the additional SAQs did not lead to a better performance in the latter. At this point we can draw several conclusions. One of them concerns the fact that generally the cohort in 2023 did not score as well overall as the one the year before. Perhaps without the guidance by the SAQs they would have scored even worse in T1. Another possibility is that the SAQs do not fulfill our hopes well enough and therefore we should leave them out for future teaching. Nevertheless, it was reported by students that they used them for preparing for the exam. Therefore, we will for future teaching still rely on this kind of support. Apparently, it seems that bonus points for the exam are the best motivation to engage into supplementary material.

A limitation of the finding reported above is the fact that the number of accesses is a first hint for the students' activities on these constructivist-oriented means, but they do not allow an insight whether cognitive processes are fostered or not. Similar applies for the importance and role of the PLGs regarding students' learning. For further investigations we suggest conducting interviews with selected students to access how cognitive processes can be supported in detail.

It should be borne in mind that any supportive measure taken with the goal of enhancing the students' learning and understanding, are merely as good as the educational reconstruction of the content. We are convinced that proper planning of the course's structure, content and at least the conceptualizations according to the MER model [1] builds up the core of the difference towards a successful course. By that, we address what is called the *deep structure* of teaching [31], a barely visible element but of great importance for students' learning. Additional and supporting the course's deep structure, we implemented elements of constructivist learning environments such as PLGs, bonus points, and SAQs, as reported above. Out of that we as lecturers and researchers do not observe that contents "had been explained dozens of times" and found that they were still "not understood" as reported in section IV. Thus, at least the lecturers job satisfaction has increased, and frustration on both sides has been largely avoided according to the student feedback reported above.

Unfortunately, we cannot provide clear evidence of an improved conceptual understanding as reported by Hake [8]. As there have been tasks (see, e.g., Ref. [32]) in the lecture, in the exercises, and for the tutorial that require rather a conceptual understanding than an elaborated calculation, we fostered the former. Lively discussions evolved around these tasks in each setting, thus supporting conceptual growth. This interesting point on improved conceptual understanding should be an issue of further investigations. We suggest a pre-post-test design (at the beginning and at the end of the semester) employing selected items from the FCI-inventory [9].

What we can derive from our data base (discussions with the student tutor and the exercise instructor, connection with the students' reports and outcomes of the exam) is a strong hint that a successful course does not rely on concessions concerning the content requirements – though they should be well-chosen. Whereas it plays a major role how students are cognitively activated, this point applies to all students in every learning setting. Moreover, it turned out to be a fruitful approach to consider how closely the content matter taught is linked to previous lectures, previous knowledge, and – most important – to the future applicability in the students' profession as teachers. The grades in the exams and the student feedback in the evaluation support the conclusion that compromises on the level and degree of sophistication of the course are the wrong way to make the lecture more effective for student learning.

Another interesting point for a future empirical evaluation of our approach is to investigate the students' mindset [33] and its development during the semester correlated to the subject matter taught. The research focus can be whether a technically demanding but profession-oriented, well-planned, and cognitively activating course supports a growth or a fixed mindset and thus fosters motivation and resilience.

Note that we speak of a course, in contrast to a lecture, which emphasizes once more, and from another perspective the interconnectedness of the physics content, concerning theory and application on problems. Similar to the teacher training studies, which we assert to be linked to future professional life, contents should be linked among each other, as well as theory and its application to meaningful problems in order to build up a coherent course. To implement this, we claim that lecture (theory part) and exercise should be of at least equal duration, ideally in the form of a lecture-exercise. Remember that in contrast we were faced with a lecture (2 h) and an exercise (1 h) for which we tried best to make a course out of it. In our view, both durations are too short, but that is another story. Still, we acknowledge that learning physics is and will stay challenging to a certain degree, which requires students' permanent effort.

Applying the MER model [1] and according to what we know from high school teaching we relied on students' prerequisites concerning mathematical skills and we based our planning on contents taught in previous lectures. There was a purposeful overlap with previous lectures concerning certain topics, which we investigated from a different perspective. This seems to be another crucial point, the validity of which surpasses local particularities related to, e.g., the specific level of instruction, the degree of mathematical sophistication or the precise legal framework under which such instruction is offered. Therefore, employing techniques known from high-school teaching for university teaching is one of the implications of our work on university teaching in general. The second implication we extract from reflective conversations is that lectures for physics education students should not be shortened versions of their counterparts for physics students. In particular, a mere shortening of the amount of content does not meet the needs of physics education students. It is a hard curriculum constraint the lecture hours of this course are less than half of those in the corresponding class at the scientific discipline of physics; nevertheless, physics education students need to get a deep understanding of the course content and not a shallow, survey-type knowledge of standard physics courses. Only in this way will they be able to answer high-school student questions properly, to explain well and to select contents forward-looking. Therefore, a new, specific format of university teaching is necessary with carefully chosen, selected contents and activities leading to engagement with conceptual questions, which is indispensable for further teaching at high schools. With our course format and accompanying peer activities, we made a concrete step towards this profession-oriented teaching format, which fits well into the model of Pedagogical Content Knowledge (PCK), as outlined up by Shulman [34], and evolved to the Refined Consensus Model (RCM) of Pedagogical Content Knowledge [35].

Lectures on the subject matter of a given discipline are meant to enhance the students' content knowledge (CK), as described by the RCM. This model outlines the various layers of knowledge and experiences that influence teachers' practices, which further on are assumed to impact student outcomes. The layers are meant to interact with and thus influence each other. The outermost one builds up the professional content bases, one element of which is the CK. The latter describes the subject matter of – in our case – physics [33]. As long as the different layers are not perceived to be independent, it will have an impact on the other layers as well, impacting how the content will be taught and the overall teaching practices at schools. Lecturers at university and their attitude towards the mere content of physics are also perceived as role models for the teacher students and *what physics is like*. We find this a strong justification for specially organized teacher courses, especially regarding the subject matter of the discipline.

On the basis of the open-ended feedback questionnaire of 2022, we gave in 2023 more breaks during the lecture for students to engage into the mathematics and link it to the underlying physics concepts. We even put more emphasis to point out to which degree certain contents may be useful for the students' further profession at school. Additionally, we tried to link as often as possible the content of the T1 course to the contents of the pedagogical seminars, thus building a bridge between the two aspects of physics and hence additionally supporting the students' PCK development.

Another feature of our approach was thorough planning and coordination among all instructors involved. We strongly recommend such practice, because it has been, in our assessment, one of the basic requirements for all persons involved to succeed, from students to lecturers. We claim that planning should always comprise foci on (a few) leading principles of content matter; selected contexts seen and described from various perspectives; and recurringly revealing their connections to the future profession as a teacher as well as to other seminars on science education research.

For the next time we will offer the course on T1 we plan to make it once more profession-oriented: First, we will identify even more situations where we point out how the content taught can be used to answer certain question which may be posed by high school students. Second, we will make students aware of the content-related rationales on which certain teaching concepts are based on. Third, whenever possible, we will make suggestions on how to link the course's content to the knowledge needed for teaching and to what degree it can be used as an additional information.

We also think that one should strive for a strong interconnection of lecture and exercises could be even stronger in the sense that acquisition of problem-solving skills may be guided more accurately. Parallel to this, one should give serious considerations exam tasks and we plan to design at least parts of them according to this focus on profession orientation. Finally, seen from the point of view of a researcher in science education research, one interesting topic to be investigated is the students' mindset and how it changes during the semester and depending on the content taught.

**Disclosure statement.** The authors report there are no competing interests to declare.

## ACKNOWLEDGEMENTS

We would especially like to thank all those who supported our project and attended the weekly team meetings and contributed to the discussions: Christian Spreitzer, instructor of the exercises and Melisa Cirak, student tutor who provided profound feedback for the lecturers. Further, we would like to express our thank to the study program managers, Kerstin Hummer, and Martin Hopf for the opportunity and freedom to conduct this collaboration. The assistance of Lukas Kolb with graphics and data analysis from Moodle is gratefully acknowledged.

## AUTHOR CONTRIBUTIONS

M.K. and C.N.L. contributed equally to this work.